\newcommand\anon[2]{{#2}} 

\documentclass[twoside,leqno,twocolumn]{article}

\usepackage[letterpaper]{geometry}

\usepackage{ltexpprt}

\usepackage{hyperref}

\usepackage{amsmath}
\usepackage{cleveref}

\usepackage{algpseudocode} 

\newtheorem{condition}{Condition}

\usepackage{physics}
\usepackage{fancyvrb}

\usepackage{bbm}
\def\cone{\mathbbm{cone}}
\def\supp{\mathbbm{supp}}

\def\cl{\mathbbm{cl}}
\def\st{\mathbbm{st}}
\def\child{\mathbbm{child}}
\def\parent{\mathbbm{parent}}

\usepackage{tikz}
\usepackage{pgflibraryshapes}
\usetikzlibrary{decorations.text}
\usetikzlibrary{decorations.pathreplacing}
\usetikzlibrary{backgrounds}
\usetikzlibrary{calc}
\usetikzlibrary{arrows}
\usetikzlibrary{snakes}
\usetikzlibrary{positioning}
\definecolor{cffffff}{RGB}{255,255,255}
\definecolor{cff00c0}{RGB}{255,0,192}
\definecolor{c0f00ff}{RGB}{15,0,255}
\definecolor{c00ff20}{RGB}{0,255,32}
\definecolor{cfffd00}{RGB}{255,253,0}
\definecolor{c00e0ff}{RGB}{0,224,255}
\usepackage{pgfplots}
\usepackage{pgfplotstable}
\graphicspath{ {./figures/} }

\usepackage{color}
\definecolor{deepblue}{rgb}{0,0,0.5}
\definecolor{deepred}{rgb}{0.6,0,0}
\definecolor{deepgreen}{rgb}{0,0.5,0}
\DeclareFixedFont{\ttb}{T1}{cmtt}{bx}{n}{10} 
\DeclareFixedFont{\ttm}{T1}{cmtt}{m}{n}{10}  
\DeclareFixedFont{\sttb}{T1}{cmtt}{bx}{n}{8} 
\DeclareFixedFont{\sttm}{T1}{cmtt}{m}{n}{8}  
\usepackage{listings}
\newcommand\cstyle{\lstset{
language=C,
basicstyle=\sttm,
otherkeywords={MPI_Comm,TS,SNES,KSP,PC,DM,Mat,Vec,VecScatter,IS,PetscSF,PetscSection,PetscObject,PetscInt,PetscScalar,PetscReal,PetscBool,InsertMode,PetscErrorCode}, 
keywordstyle=\sttb\color{deepblue},
emph={PETSC_COMM_WORLD,PETSC_NULL,SNES_NGMRES_RESTART_PERIODIC},          
emphstyle=\sttb\color{deepred},    
commentstyle=\sttm\color{brown},
stringstyle=\sttm\color{deepgreen},
frame=tb,                         
showstringspaces=false,           %
columns=fullflexible              
}}

\newcommand\cinlinestyle{\lstset{
language=C,
basicstyle=\ttm,
otherkeywords={MPI_Comm,TS,SNES,KSP,PC,DM,Mat,Vec,VecScatter,IS,PetscSF,PetscSection,PetscObject,PetscInt,PetscScalar,PetscReal,PetscBool,InsertMode,PetscErrorCode}, 
keywordstyle=\ttb\color{deepblue},
emph={PETSC_COMM_WORLD,NULL,SNES_NGMRES_RESTART_PERIODIC},          
emphstyle=\ttb\color{deepred},    
commentstyle=\ttm\color{brown},
stringstyle=\ttm\color{deepgreen},
frame=tb,                         
showstringspaces=false,           %
columns=fullflexible              
}}

\lstnewenvironment{cprog}[1][]
{
\cstyle
\lstset{#1}
}
{}

\newcommand\cinline[1]{{\cinlinestyle\lstinline!#1!}}

\begin{document}

\newcommand\relatedversion{}
\renewcommand\relatedversion{\thanks{The full version of the paper can be accessed at \anon{\protect\url{redacted-url}}{\protect\url{https://arxiv.org/abs/1902.09310}}}} 

\title{\Large Transformations of Computational Meshes}
\anon{
    \author{Submission ID XYZ}  
}
{ 
    \author{Matthew G. Knepley\thanks{University at Buffalo}}
}

\date{}

\maketitle




\fancyfoot[R]{\scriptsize{Copyright \textcopyright\ 2024\\
Copyright for this paper is retained by authors}}



\begin{abstract} \small\baselineskip=9pt Computational meshes, as a way to partition space, form the basis of much of PDE simulation technology, for instance for the finite element and finite volume discretization methods. In complex simulations, we are often driven to modify an input mesh, for example, to refine, coarsen, extrude, change cell types, or filter it. Mesh manipulation code can be voluminous, error-prone, spread over many special cases, and hard to understand and maintain by subsequent developers. We present a simple, table-driven paradigm for mesh transformation which can execute a large variety of transformations in a performant, parallel manner, along with experiments in the open source library PETSc which can be run by the reader.
\end{abstract}

\section{Introduction.}

In PDE simulations, computational meshes are tools for partitioning the domain of the PDE, allowing operations such as integration to be carried out separately on each piece. This is the heart of both the finite element and finite volume methods and their many derivatives and specializations. For complex simulations, these meshes are often created using mesh generators~\cite{shewchuk96,blackerbohnhoffedwards1994,si2015} or constructed from CAD~\cite{dentonthesis2022}. However, in complex simulations, it is often necessary to alter these meshes based upon the input parameters, solution characteristics, or hardware environment. For example, we could refine or coarsen regions of the mesh, create missing cells of dimension $k$ ($k$-cells) such as faces or edges (mesh interpolation), change cell types~\cite{timalsinaknepley2023}, filter out portions of the mesh, or extrude new $k$-cells from existing faces. These capabilities are often absent from mesh generators, difficult to access from running simulations due to the lack of an API, only partially present in special-purpose libraries, and greatly complicated user code.

We present a simple theory of mesh transformations based upon transformations of an underlying graph, isomorphic to the mesh. We can construct simple, tabular descriptions of these graph transformations which correspond to the common mesh transformation listed above. In this way, we can write a single routine to apply the transformation, independent of the precise nature of the transformation. This routine can be made performant, and has also been parallelized, so that all transformations run scalably.

We have implemented this algorithm in the PETSc library~\cite{petsc-user-ref,petsc-web-page} for scientific computing, as part of the DMPlex module~\cite{langemitchellknepleygorman2015,knepleylangegorman2017} which supports unstructured meshes. DMPlex has support for unstructured meshes of arbitrary dimension and cell shape, large scale parallelism~\cite{knepleylangegorman2017,parsani2021}, parallel I/O and checkpointing~\cite{haplaknepleyafanasievboehmdrielkrischerfichtner2020,hamhaplaknepleymitchellsagiyama2024}, and CAD interfacing~\cite{dentonthesis2022}. We illustrate the functionality with simple examples from PETSc that can be run by the user.

\subsection{Meshes as Graphs.}

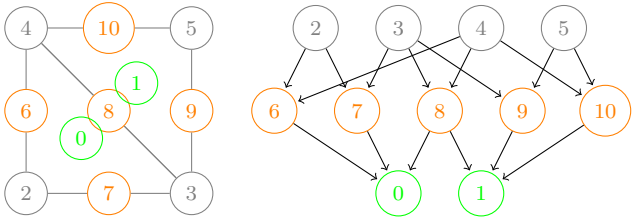
\begin{figure}
\newcommand{\vStart}{2}
\newcommand{\vEnd}{5}
\newcommand{\numVertices}{4}
\newcommand{\vShift}{3.50}
\newcommand{\eStart}{6}
\newcommand{\eEnd}{10}
\newcommand{\eShift}{3.00}
\newcommand{\numEdges}{5}
\newcommand{\cStart}{0}
\newcommand{\cEnd}{1}
\newcommand{\numCells}{2}
\newcommand{\cShift}{4.50}
\begin{tikzpicture}[scale = 1.1,font=\fontsize{8}{8}\selectfont]
\path (0.,0.) node(2_0) [draw,shape=circle,color=gray] {2};
\path (2.,0.) node(3_0) [draw,shape=circle,color=gray] {3};
\path (0.,2.) node(4_0) [draw,shape=circle,color=gray] {4};
\path (2.,2.) node(5_0) [draw,shape=circle,color=gray] {5};
\path
(0.,1.) node(6_0) [draw,shape=circle,color=orange] {6} --
(1.,0.) node(7_0) [draw,shape=circle,color=orange] {7} --
(1.,1.) node(8_0) [draw,shape=circle,color=orange] {8} --
(2.,1.) node(9_0) [draw,shape=circle,color=orange] {9} --
(1.,2.) node(10_0) [draw,shape=circle,color=orange] {10} --
(0,0);
\draw[color=gray] (4_0) -- (6_0) -- (2_0) -- (7_0) -- (3_0) -- (8_0) -- (4_0);
\draw[color=gray] (3_0) -- (9_0) -- (5_0) -- (10_0) -- (4_0) -- (8_0) -- (3_0);
\path (0.666667,0.666667) node(0_0) [draw,shape=circle,color=green] {0};
\path (1.33333,1.33333) node(1_0) [draw,shape=circle,color=green] {1};
\foreach \c in {\cStart,...,\cEnd}
{
  \node(\c_0) [draw,shape=circle,color=green,minimum size = 6mm] at (\cShift+\c-\cStart,0) {\c};
}
\foreach \e in {\eStart,...,\eEnd}
{
  \node(\e_0) [draw,shape=circle,color=orange,minimum size = 6mm] at (\eShift+\e-\eStart,1) {\e};
}
\foreach \v in {\vStart,...,\vEnd}
{
  \node(\v_0) [draw,shape=circle,color=gray,minimum size = 6mm] at (\vShift+\v-\vStart,2) {\v};
}
\draw[->, shorten >=1pt] (6_0) -- (0_0);
\draw[->, shorten >=1pt] (7_0) -- (0_0);
\draw[->, shorten >=1pt] (8_0) -- (0_0);
\draw[->, shorten >=1pt] (9_0) -- (1_0);
\draw[->, shorten >=1pt] (10_0) -- (1_0);
\draw[->, shorten >=1pt] (8_0) -- (1_0);
\draw[->, shorten >=1pt] (4_0) -- (6_0);
\draw[->, shorten >=1pt] (2_0) -- (6_0);
\draw[->, shorten >=1pt] (2_0) -- (7_0);
\draw[->, shorten >=1pt] (3_0) -- (7_0);
\draw[->, shorten >=1pt] (3_0) -- (8_0);
\draw[->, shorten >=1pt] (4_0) -- (8_0);
\draw[->, shorten >=1pt] (3_0) -- (9_0);
\draw[->, shorten >=1pt] (5_0) -- (9_0);
\draw[->, shorten >=1pt] (5_0) -- (10_0);
\draw[->, shorten >=1pt] (4_0) -- (10_0);
\end{tikzpicture}
  \caption{Two adjacent triangles, and the corresponding Hasse diagram.\label{fig:triangleDoublet}}
\end{figure}

It was recognized early on in the study of topology that cellular decompositions of space could be represented as graphs~\cite{aleksandrov98}. We illustrate the idea with the simple mesh shown in \cref{fig:triangleDoublet}, which can be created by PETSc using
\begin{Verbatim}[fontsize=\small,commandchars=\\\{\}]
  \$PETSC_DIR/src/dm/impls/plex/tests/ex1
   -dm_plex_box_faces 1,1 -dm_view ::ascii_latex
   -dm_plex_view_hasse -dm_plex_view_tikzscale 1.5
\end{Verbatim}
The graph associated to a mesh, called a Hasse Diagram~\cite{hassediagram}, has a vertex for every $k$-cell in the graph. In our case that means four vertices for 0-cells (mesh vertices) in the top layer, five vertices for 1-cells (mesh edges), and two vertices for 2-cells (mesh triangles). The graph edges represent the adjacency, or boundary, relation. For example, edges 6, 7, and 8 form the boundary of triangle 0, and thus there is a graph edge from graph vertices 6, 7, and 8 to vertex 0. Likewise, there are graph edges from vertices 2 and 4 to vertex 6, representing the two mesh vertices bounding edge 6. Clearly, this correspondence can be established independent of the shape of each $k$-cell and independent of the dimension of the mesh, and it is isomorphic to the original mesh. Also note that the Hasse Diagram is a directed acyclic graph (DAG) which is stratified into layers by the dimension $k$ of each vertex. Thus the graph forms a graded poset, ordered by the adjacency relation.

Using language adapted from combinatorial topology~\cite{knepleykarpeev05,knepleykarpeev09}, we refer to the set of graph vertices connected to in-edges as the \textit{cone} of a vertex, and the dual set of vertices connected to out-edges as the \textit{support}. The transitive closure of the in-edge relation is called the \textit{closure} of a vertex, and its dual on out-edges is the \textit{star}.

Let $P_k(p, q)$ be the relation ``there exists a directed path of length $k$ from $p$ to $q$ in the Hasse diagram'', and let $P$ be the union of all these relations
\begin{align}
  P = \bigcup_k P_k.
\end{align}
We can define the cone and support operations as a duality relation
\begin{align}\label{eq:coneDuality}
  q \in \cone(p) \iff P_1(p, q) \iff p \in \supp(q).
\end{align}
And we can generalize this to transitive closures,
\begin{align}\label{eq:closureDuality}
  q \in \cl(p) \iff P(p, q) \iff p \in \st(q).
\end{align}

We need a final ingredient to fully specify a mesh, namely an \textit{orientation} for each graph edge (Hasse Diagram arrow) attaching a $(k-1)$-cell to a $k$-cell. Without orientations we could not recover boundaries as oriented manifolds, which for example would prevent us from calculating correct normals. In addition, we would like to preserve the property that the boundary operator is nilpotent. We also would like a canonical order on the closure of each cell, for which we make use of the orientation information.

Each $(k-1)$-cell is stored in some canonical order. The orientation describes how to rotate/reflect it before attaching it to a particular $k$-cell. We define an orientation as an integer labeling an element of the dihedral group for the given $(k-1)$-cell, which is the group of symmetries for the cell. For example, vertices have a single orientation, labeled 0. The the dihedral group of a segment (a 1-cell) is just $F_2$, so it can attach in the canonical order or reversed, and we use 0 and -1 to indicate the orientation. The highest dimensional cells are not attached to anything, and thus always use the canonical (stored) order, which we term the identity orientation. The dihedral group of the triangle is shown in~\cref{tab:triOrient}. We can think of these orientations as labels for the cone edges in our Hasse Diagram.

In the same way that a mesh can be defined by specifying the boundary of each $k$-cell, we can specify the Hasse Diagram by defining the cone of each graph vertex. This is, in fact, how the DMPlex module in the PETSc library manages unstructured meshes. All mesh operations are translated to equivalent graph operations. When representing parallel meshes, graphs representing the local mesh are held on each process, and then graph vertices are identified between processes using the PETSc SF object~\cite{petscsf2022}, which is merely a biparatite graph connecting the vertex pairs.

\section{A Transformation Grammar.}\label{sec:grammar}

The process of transforming one mesh into another can be reexpressed as the transformation of one graph into another. In order to create the new transformed mesh graph, we apply production rules to the original graph. The special feature of our rules is that they produce both a new graph vertex and its cone with oriented edges, rather than just the vertex itself~\cite{courcelleengelfriet2012}. The cone is composed of points produced by other rules, and thus we must have a way of refering to all points that are produced from a given graph.

Suppose that a point $p$ in the input mesh produces a point $q$ in the transformed mesh, and we call $q$ the \textit{child} of $p$. We would like a kind of locality, meaning that the cone of $q$ is easily discoverable given $p$. The simplest rule of this kind would be that the cone of $p$ produces the cone of $q$. However, this rules out many transformations, such as regular refinement of a mesh. Thus we begin with a slightly more expansive rule
\begin{condition}\label{cond:local}
  The cone of a point $q$ of the transformed mesh is produced by the closure of a point $p$ in the input mesh.
\end{condition}
which we can write
\begin{align}
  \cone(\child(p)) \in \child(\cl(p)),
\end{align}
where we use $\child(p)$ to indicate the set of points produced by point $p$, or the ``production cone'' of $p$. There is a similar dual notion, a ``production support'', which we call $\parent(q)$ meaning the set of points which can produce $q$. With this condition, we need only compute the part of the transformed mesh produced by the closure of $p$ in order to capture the cone of $q$. This gives us an easy way to bound both computation and storage costs for our virtual mesh. By \textit{virtual}, we mean a mesh that is not concretely stored, but rather it is partially created on the fly as needed to answer queries, using the original mesh and description of the transformation.

What about the closure of $q$? Consider a point $q'$ in the cone of $q$, so that
\begin{align}
  q' &\in \cone(q) \\
     &\in \child(\cl(p))
\end{align}
by Condition~\ref{cond:local}. This means there is some point $p'$ in the closure of $p$ that produces $q'$. Thus,
\begin{align}
  \cone(q') &\in \cone(\child(p')) \\
            &\in \child(\cl(p')) \\
            &\in \child(\cl(p))
\end{align}
where the last line follows because transitive closures are nested. By reasoning this way for each point in the closure, we can conclude that
\begin{align}\label{eq:childClosure}
  \cl(\child(p)) &\in \child(\cl(p)).
\end{align}

We have now bounded the complexity to compute our transformed mesh, but how costly is it to identify the produced points? In concrete terms, how can one compute a total order on the points in the transformed mesh? Consider the situation for mesh interpolation. We can think of each cell producing the faces and edges in its closure. However, this would mean that multiple cells would produce the same face or edge. To compute a numbering, we have to compute a signature for each point introduced, say its vertex cone, and then compare signatures to establish identity. In order to limit the resources for this comparison, we need some limit on the parent set for a point $q$. The simplest condition is
\begin{condition}\label{cond:unique}
  A point $q$ of the transformed mesh is produced by only one $p$ in the input mesh, such that $|\parent(q)| = 1$.
\end{condition}
This rule allows a unique numbering to be calculated given only a prior numbering of the producing graph. Even in parallel, a local numbering can be calculated independently and patched together using PetscSF. Using Condition~\ref{cond:unique}, we can now bound the cost of support queries. Let a point $q$ be produced by a point $p$, and consider a point $q'$ in the star of $q$,
\begin{align*}
  q' \in \st(q) \iff q \in \cl(q')
\end{align*}
from Eq.~\ref{eq:closureDuality}. Let $q'$ be produced by a point $p'$,
\begin{align}
  q' \in \child(p')
\end{align}
so that
\begin{align}
  \cl(q') &\in \child(\cl(p')) \\
  q       &\in \child(\cl(p'))
\end{align}
We can now use that fact that parents are unique, Condition~\ref{cond:unique},
\begin{align}
  \parent(q) &\in \cl(p') \\
  p          &\in \cl(p') \\
  p'         &\in \st(p)
\end{align}
so that we have shown
\begin{align}\label{eq:childStar}
  \st(\child(p)) \in \child(\st(p)).
\end{align}
Thus, to compute a star for any point in the transformed mesh, we need only consider the star of the parent in the input mesh.

A more sophisticated condition would bound the set of possible parents. For example, we could require that
\begin{align}
  \parent(\st(q)) \in \st(\parent(q))
\end{align}
which would produce the same kind of locality. This is satisfied by the interpolation algorithm, even though parents are not unique. In parallel, this could become problematic as regions to check spread across process boundaries. Interpolation, fortunately, has another property. All the produced points that are shared are guaranteed to have cones which are also shared, but we have not yet generalized this property to our set of transformations.

\subsection{Definition}\label{sec:transformDef}

A transformation is be defined by its action on each $k$-cell, in that for each $k$-cell in the source mesh the transformation produces a set of $l$-cells (which can be the empty set) in the target mesh. In the simplest examples, such as regular refinement, the action depends only on the cell type. However, we allow the transformation to make different decisions for cells of the same type. In the implementation, this is accomplished using a label to differentiate the cells, giving each cell a \textit{transformation type} to refine its cell type. Thus, in the discussion below we refer to transformation type, but the reader can imagine this as a stand-in for the cell type in order to get an intuitive feel for the algorithm. As we discuss the definition below, we use regular tetrahedral refinement as a non-trivial example to illustrate the stages. We show excerpts of library code in order to make our description concrete and precise, however users would not have to write this kind of code unless they wanted to define a new transformation type.

In our definition, we first indicate which cell types are produced for a given transformation type. This simplifies the description by separating the different transformation rules. It also allows the stratification of the target mesh to be easily computed. By stratification, we meanthe separation of the graph into layers by cell dimension, since there are only edges between adjacent dimensions. In fact, this gives the Hasse diagram the structure of a graded poset. In our tetrahedral refinement example, we need to consider the transformation of four cell types: vertices, edges, triangular faces, and tetrahedral cells. The vertices produce identical copies, and thus have a single production type {\sc point}. Edges are split into two pieces, yielding one {\sc point} in the center, and two {\sc segments}. The triangular faces are divided into four, producing three {\sc segments} and four {\sc triangles}. Notice that the subdivided triangles do not count the edges and vertices introduced on the boundary because those are handled by the edge transformation rule. In general, the transformation rule only deals with the interior, not the boundary of a cell. Finally, the tetrahedron is divided into eight, producing one {\sc segment}, eight {\sc triangles}, and eight {\sc tetrahedra}. Since several points can be produced with the same cell type, we number them using a \textit{replica number}, which we denote as $r$.

In order to describe the cells which are produced, we need to know their cones and cone orientations. The cones consist of points in the target mesh, so we must have some way of identifying them. If we just needed to refer to points produced by the current point, we could use the celltype and replica number. However, this is not in general sufficient. Instead we make use of Condition~\ref{cond:local}, which says that the cone of any point produced must be contained in the set of points produced from the closure of the original point. Thus, we need to first locate a point in the closure of the original point, and then specify the target point produced from it.

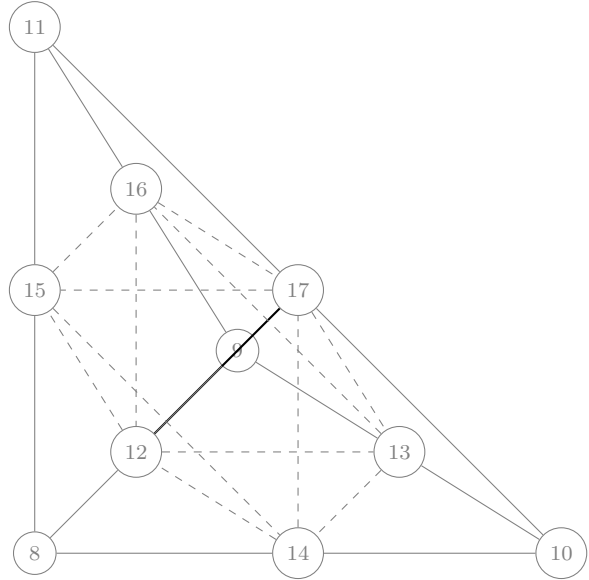
\begin{figure}
\begin{tikzpicture}[scale = 1.75,font=\fontsize{8}{8}\selectfont]
\path (-2.,-2.,2.) node(8_0) [draw,shape=circle,color=gray] {8};
\path (-2.,-2.,-2.) node(9_0) [draw,shape=circle,color=gray] {9};
\path (2.,-2.,2.) node(10_0) [draw,shape=circle,color=gray] {10};
\path (-2.,2.,2.) node(11_0) [draw,shape=circle,color=gray] {11};
\path (-2.,-2.,-0.) node(12_0) [draw,shape=circle,color=gray] {12};
\path (0.,-2.,-0.) node(13_0) [draw,shape=circle,color=gray] {13};
\path (0.,-2.,2.) node(14_0) [draw,shape=circle,color=gray] {14};
\path (-2.,0.,2.) node(15_0) [draw,shape=circle,color=gray] {15};
\path (-2.,0.,-0.) node(16_0) [draw,shape=circle,color=gray] {16};
\path (0.,0.,2.) node(17_0) [draw,shape=circle,color=gray] {17};
\draw[color=black,thick] (12_0) -- (17_0);
\draw[color=gray,dashed] (12_0) -- (13_0);
\draw[color=gray,dashed] (13_0) -- (14_0);
\draw[color=gray,dashed] (14_0) -- (12_0);
\draw[color=gray,dashed] (15_0) -- (16_0);
\draw[color=gray,dashed] (16_0) -- (12_0);
\draw[color=gray,dashed] (12_0) -- (15_0);
\draw[color=gray,dashed] (14_0) -- (17_0);
\draw[color=gray,dashed] (17_0) -- (15_0);
\draw[color=gray,dashed] (15_0) -- (14_0);
\draw[color=gray,dashed] (13_0) -- (16_0);
\draw[color=gray,dashed] (16_0) -- (17_0);
\draw[color=gray,dashed] (17_0) -- (13_0);
\draw[color=gray] (8_0) -- (12_0);
\draw[color=gray] (12_0) -- (9_0);
\draw[color=gray] (9_0) -- (13_0);
\draw[color=gray] (13_0) -- (10_0);
\draw[color=gray] (10_0) -- (14_0);
\draw[color=gray] (14_0) -- (8_0);
\draw[color=gray] (8_0) -- (15_0);
\draw[color=gray] (15_0) -- (11_0);
\draw[color=gray] (11_0) -- (16_0);
\draw[color=gray] (16_0) -- (9_0);
\draw[color=gray] (10_0) -- (17_0);
\draw[color=gray] (17_0) -- (11_0);
\path (-1.5,-1.5,1.5) node(0_0) [] {};
\path (-1.5,-1.5,-0.5) node(1_0) [] {};
\path (0.5,-1.5,1.5) node(2_0) [] {};
\path (-1.5,0.5,1.5) node(3_0) [] {};
\path (-1.,-1.,1.5) node(4_0) [] {};
\path (-1.,-1.,0.5) node(5_0) [] {};
\path (-0.5,-1.5,1.) node(6_0) [] {};
\path (-1.5,-0.5,1.) node(7_0) [] {};
\end{tikzpicture}
  \caption{A single regular refinement of a tetrahedron. We have indicated the interior segment in black, and used dashed lines for the segments produced by the boundary faces.\label{fig:tetRefReg}}
\end{figure}

We use the tetrahedron as an example, shown in Fig.~\ref{fig:tetRefReg}, as it is the simplest shape with non-trivial cone production. A single parent tetrahedron is split into eight child tetrahedra. In the figure, the child tetrahedra are numbered 0--7 (not shown), and the vertices produced are numbered 8--17, as this is how the Hasse diagram is stored in memory.

Let us first consider the lone segment produced by dividing the tetrahedron. We can describe the cone of this segment with the following array
\begin{Verbatim}[fontsize=\small]
  {DM_POLYTOPE_POINT, 2, 0, 0, 0,
   DM_POLYTOPE_POINT, 2, 2, 1, 0};
\end{Verbatim}
The first cone point is a vertex, and we arrive at the producing point by taking two cones. First, we take cone point 0 of the tetrahedron, which is the bottom face. Then we take cone point 0 in that bottom face, which is the left edge. Finally, we take the first vertex produced by that edge, which is replica 0. In the reference cell, this vertex has coordinates $(-1, 0, -1)$. We can carry out the same computation for the second cone point, which also requires taking two cones. We take cone point 2 of the tetrahedron, which is the front face, and then cone point 1 of that face, which is the diagonal edge, and finally the first vertex produced by that edge, which has coordinates $(0, -1, 0)$. To complete the definition, we should also specify the orientation of each cone point, but vertices can only have orientation 0. Notice that the description of these points is not unique. For example, we could have used face 3 to extract the second vertex.

We proceed in the same manner for all target points produced from this source point. The first internal triangular subface can be described using
\begin{Verbatim}[fontsize=\small]
  {DM_POLYTOPE_SEGMENT, 1, 0, 2,
   DM_POLYTOPE_SEGMENT, 1, 1, 2,
   DM_POLYTOPE_SEGMENT, 1, 2, 2};
\end{Verbatim}
The first edge is produced from a point in the cone of the tetrahedron, so we only need 1 cone operation. It comes from face 0 of the tetrahedron, which is the bottom face, and is the third segment produced, which means it connects the left edge with the front edge. The second edge is produced from face 1, which is the left face, and is the third segment produced, meaning it connects the bottom edge (left edge from before) with the vertical edge. Finally, the third edge is produced from face 2, which is the front face, and is the third segment produced, which connects the vertical edge to the bottom edge (front edge from the first face). Thus we have a triangle, and the edges are properly oriented. On other internal triangles, for example
\begin{Verbatim}[fontsize=\small]
  {DM_POLYTOPE_SEGMENT, 1, 0, 2,
   DM_POLYTOPE_SEGMENT, 0,    0,
   DM_POLYTOPE_SEGMENT, 1, 2, 0};
\end{Verbatim}
some segments need to be reversed, in this case the last one. Therefore we must also provide an orientation array, which here would look like
\begin{Verbatim}[fontsize=\small]
  {0, 0, -1};
\end{Verbatim}
Note also that the middle segment is produced directly by the tetrahedron, and thus 0 cones need to be taken.

If we look at the first subtetrahedron, we have the expression
\begin{Verbatim}[fontsize=\small]
  {DM_POLYTOPE_TRIANGLE, 1, 0, 0,
   DM_POLYTOPE_TRIANGLE, 1, 1, 0,
   DM_POLYTOPE_TRIANGLE, 1, 2, 0,
   DM_POLYTOPE_TRIANGLE, 0,    0};
\end{Verbatim}
The first face is produced from face 0 in the tetrahedron, which is the bottom face. It is the first triangle produced, meaning it is the triangle containing the lower-left vertex. The second face is produced from face 1, and is also the triangle containing the lower-left vertex. Similarly, the third face comes from face 2 of the tetrahedron, and is the triangle containing the lower-left vertex. Finally, the last face is produced directly by the tetrahedron itself, and is the first such face, which divides the lower-left vertex, $(-1, -1, -1)$, from the rest of the tetrahedron. Also, each of these faces is properly oriented. However, this is not true of all the subtetrahedra, so we need to provide an orientation array for the faces.

This same procedure works for quadrilaterals, hexahedra, and pyramids in PETSc, and we can imagine defining tables for any $k$-cell which can be included in the mesh. The tables can be seen in the PETSc source code in the \texttt{plexrefregular.c} file.

Finally, we need to know how the orientation of a child point behaves as a function of the orientation of the parent point. In fact, as the orientation of the parent changes, replicas may change places, so we need to also know the new replica number. This again can be encoded in a small table for each transform type. As an example, consider the four child triangles produced by a parent triangle. \Cref{tab:triOrient} shows the possible orientations and corresponding arrangements for a triangle. A triangle orientation $o \in [-3, 2]$, where the negative orientations imply reflection. For each triangle orientation, we tabulate the replica permutation and orientation transformation. The array is as follows,
\begin{cprog}
  a =
  {1, -3,  0, -3,  2, -3,  3, -2,
   0, -2,  2, -2,  1, -2,  3, -1,
   2, -1,  1, -1,  0, -1,  3, -3,
   0,  0,  1,  0,  2,  0,  3,  0,
   1,  1,  2,  1,  0,  1,  3,  1,
   2,  2,  0,  2,  1,  2,  3,  2};
\end{cprog}
so that for parent orientation $o_p$, we get child replica number $r_c$ to substitute for replica $r$ with transformation $o_c$ using
\begin{align}
  r_c &= a[(o_p + 3) * 8 + r * 2] \\
  o_c &= a[(o_p + 3) * 8 + r * 2 + 1]
\end{align}
The final orientation of the child point is the composition of its given orientation $o$ with $o_c$. For example, the identity transformation, orientation 0, shows that $r_c = r$ and $o_c = 0$. If we rotate the triangle $60^\circ$ clockwise, orientation 1, then the three triangles at the corners switch places, the permutation $[1\ 2\ 0]$, and each one also rotates $60^\circ$ clockwise, orientation 1, while the central triangle remains but with orientation 1 as well. If we interchange the second and third vertices, orientation -2, then the second and third child triangles interchange, permutation $[0\ 2\ 1]$, and each has orientation -2, and the central triangle again remains fixed but with orientation -1, since it had its first and third vertices interchanged instead.

\begin{table}
\begin{tabular}{lccc}
Ornt & 0 & 1 & 2 \\
\vbox{\hbox{Arr}\hbox{\rule{0pt}{6ex}}} &
\begin{tikzpicture}[scale = 0.4,font=\fontsize{8}{8}\selectfont]
\path (-2.,-2.) node(1_0) [draw,shape=circle,color=gray] {1};
\path (2.,-2.) node(2_0) [draw,shape=circle,color=gray] {2};
\path (-2.,2.) node(3_0) [draw,shape=circle,color=gray] {3};
\path
(0.,-2.) node(4_0) [draw,shape=circle,color=gray] {4} --
(0.,0.) node(5_0) [draw,shape=circle,color=gray] {5} --
(-2.,0.) node(6_0) [draw,shape=circle,color=gray] {6} --
(0,0);
\draw[color=gray] (1_0) -- (4_0) -- (2_0) -- (5_0) -- (3_0) -- (6_0) -- (1_0);
\path (-0.666667,-0.666667) node(0_0) [] {};
\end{tikzpicture}
&
\begin{tikzpicture}[scale = 0.4,font=\fontsize{8}{8}\selectfont]
\path (-2.,2.) node(1_0) [draw,shape=circle,color=gray] {1};
\path (-2.,-2.) node(2_0) [draw,shape=circle,color=gray] {2};
\path (2.,-2.) node(3_0) [draw,shape=circle,color=gray] {3};
\path
(-2.,0.) node(4_0) [draw,shape=circle,color=gray] {4} --
(0.,-2.) node(5_0) [draw,shape=circle,color=gray] {5} --
(0.,0.) node(6_0) [draw,shape=circle,color=gray] {6} --
(0,0);
\draw[color=gray] (2_0) -- (5_0) -- (3_0) -- (6_0) -- (1_0) -- (4_0) -- (2_0);
\path (-0.666667,-0.666667) node(0_0) [] {};
\end{tikzpicture}
&
\begin{tikzpicture}[scale = 0.4,font=\fontsize{8}{8}\selectfont]
\path (2.,-2.) node(1_0) [draw,shape=circle,color=gray] {1};
\path (-2.,2.) node(2_0) [draw,shape=circle,color=gray] {2};
\path (-2.,-2.) node(3_0) [draw,shape=circle,color=gray] {3};
\path
(0.,0.) node(4_0) [draw,shape=circle,color=gray] {4} --
(-2.,0.) node(5_0) [draw,shape=circle,color=gray] {5} --
(0.,-2.) node(6_0) [draw,shape=circle,color=gray] {6} --
(0,0);
\draw[color=gray] (3_0) -- (6_0) -- (1_0) -- (4_0) -- (2_0) -- (5_0) -- (3_0);
\path (-0.666667,-0.666667) node(0_0) [] {};
\end{tikzpicture}
\\
\hline
\hline
Ornt & -1 & -2 & -3 \\
\vbox{\hbox{Arr}\hbox{\rule{0pt}{6ex}}} &
\begin{tikzpicture}[scale = 0.4,font=\fontsize{8}{8}\selectfont]
\path (-2.,2.) node(1_0) [draw,shape=circle,color=gray] {1};
\path (2.,-2.) node(2_0) [draw,shape=circle,color=gray] {2};
\path (-2.,-2.) node(3_0) [draw,shape=circle,color=gray] {3};
\path
(0.,0.) node(4_0) [draw,shape=circle,color=gray] {4} --
(0.,-2.) node(5_0) [draw,shape=circle,color=gray] {5} --
(-2.,0.) node(6_0) [draw,shape=circle,color=gray] {6} --
(0,0);
\draw[color=gray] (3_0) -- (5_0) -- (2_0) -- (4_0) -- (1_0) -- (6_0) -- (3_0);
\path (-0.666667,-0.666667) node(0_0) [] {};
\end{tikzpicture}
&
\begin{tikzpicture}[scale = 0.4,font=\fontsize{8}{8}\selectfont]
\path (-2.,-2.) node(1_0) [draw,shape=circle,color=gray] {1};
\path (-2.,2.) node(2_0) [draw,shape=circle,color=gray] {2};
\path (2.,-2.) node(3_0) [draw,shape=circle,color=gray] {3};
\path
(-2.,0.) node(4_0) [draw,shape=circle,color=gray] {4} --
(0.,0.) node(5_0) [draw,shape=circle,color=gray] {5} --
(0.,-2.) node(6_0) [draw,shape=circle,color=gray] {6} --
(0,0);
\draw[color=gray] (1_0) -- (6_0) -- (3_0) -- (5_0) -- (2_0) -- (4_0) -- (1_0);
\path (-0.666667,-0.666667) node(0_0) [] {};
\end{tikzpicture}
&
\begin{tikzpicture}[scale = 0.4,font=\fontsize{8}{8}\selectfont]
\path (2.,-2.) node(1_0) [draw,shape=circle,color=gray] {1};
\path (-2.,-2.) node(2_0) [draw,shape=circle,color=gray] {2};
\path (-2.,2.) node(3_0) [draw,shape=circle,color=gray] {3};
\path
(0.,-2.) node(4_0) [draw,shape=circle,color=gray] {4} --
(-2.,0.) node(5_0) [draw,shape=circle,color=gray] {5} --
(0.,0.) node(6_0) [draw,shape=circle,color=gray] {6} --
(0,0);
\draw[color=gray] (2_0) -- (4_0) -- (1_0) -- (6_0) -- (3_0) -- (5_0) -- (2_0);
\path (-0.666667,-0.666667) node(0_0) [] {};
\end{tikzpicture}
\end{tabular}
\caption{The dihedral group $D_3$ for the triangle~\label{tab:triOrient}.}
\end{table}

Note here that the combination of a base mesh and a transformation could be said to constitute a new mesh, in that it can itself satisfy all mesh queries for the transformed mesh. There is no need to concretely instantiate the mesh itself. This is elaborated upon in~\cref{sec:ephemeral}.

\subsection{Numbering.}

The defining feature of our formalization of a mesh transformation is that the transformed mesh can be known efficiently purely from the definition of the input mesh and the transformation. The first step is to introduce a map from source points, those in the original mesh, to target points, those in the transformed mesh, and also its inverse. We can think of this numbering as an index structure, built on top of the transformation mechanism discussed above.

Suppose that we are given a source point $p$ and its transformation type $t_p$, a cell type that is being produced $t_q$, and a replica number $r$, then we can compute the number of the produced point $q$ in the transformed mesh. This only needs to be a bijective map to the integers, but we further require that it map to a contiguous set of integers to facilitate fast indexing in the implementation.

In our implementation, we compute this numbering by constructing a table of offsets. During the setup phase, we precompute data needed for the index. First, a total order on transformation/cell types is defined. This might not match the enumeration value since we often want to number cells, then vertices, then faces, and then edges so that mesh interpolation does not affect the ordering. Next, we run over the original mesh, consulting our production rule for each point, and record the outputs. This allows us to calculate offsets for each transformation type in the transformed mesh. In fact, we calculate the offset for the first cell of given child cell type produced by a cell of given parent cell type. Thus given the parent transformation type, we can lookup the offset for the child cell type, which we call $o(t_p, t_q)$.

Next we look at the description for the given transform type. We can see the number of replicas of the child cell type for each parent point, called $N_r(t_p)$, which we multiply by the reduced point number $rp$, meaning that the parent point is the $rp$th point of the parent transformation type. In total, our child point number is
\begin{align}
  q = o(t_p, t_q) + rp \times N_r(t_p) + r.
\end{align}
In parallel, we must communicate these offsets to all processes so that remote point numbers can be calculated for newly introduced members of the transformed PetscSF. However, this is a very small amount of data that can be sent with a single allreduce.

\section{Extrusion}

Extrusion is the transformation of a mesh to one of higher dimension replacing each cell by a new cell that has two copies of the original as faces, using a tensor product construction. For example, an extruded segment would become a quadrilateral, and an extruded triangle becomes a triangular prism. Below, we detail this transformation and the many customizations one can make.

As a simple example, we create a quadrilateral mesh of the unit square,
\begin{Verbatim}[fontsize=\small]
  ./ex1 -dm_plex_simplex 0 -dm_plex_box_faces 3,3
    -dm_view :mesh.tex:ascii_latex
    -dm_plex_view_numbers 0
    -dm_plex_view_colors_depth 1,0,0
\end{Verbatim}
producing the simple mesh below.
\begin{center}
\begin{tikzpicture}[scale = 1.,font=\fontsize{8}{8}\selectfont]
\path (0.,0.) node(9_0) [fill,inner sep=1pt,shape=circle,color=gray] {};
\path (0.666667,0.) node(10_0) [fill,inner sep=1pt,shape=circle,color=gray] {};
\path (1.33333,0.) node(11_0) [fill,inner sep=1pt,shape=circle,color=gray] {};
\path (2.,0.) node(12_0) [fill,inner sep=1pt,shape=circle,color=gray] {};
\path (0.,0.666667) node(13_0) [fill,inner sep=1pt,shape=circle,color=gray] {};
\path (0.666667,0.666667) node(14_0) [fill,inner sep=1pt,shape=circle,color=gray] {};
\path (1.33333,0.666667) node(15_0) [fill,inner sep=1pt,shape=circle,color=gray] {};
\path (2.,0.666667) node(16_0) [fill,inner sep=1pt,shape=circle,color=gray] {};
\path (0.,1.33333) node(17_0) [fill,inner sep=1pt,shape=circle,color=gray] {};
\path (0.666667,1.33333) node(18_0) [fill,inner sep=1pt,shape=circle,color=gray] {};
\path (1.33333,1.33333) node(19_0) [fill,inner sep=1pt,shape=circle,color=gray] {};
\path (2.,1.33333) node(20_0) [fill,inner sep=1pt,shape=circle,color=gray] {};
\path (0.,2.) node(21_0) [fill,inner sep=1pt,shape=circle,color=gray] {};
\path (0.666667,2.) node(22_0) [fill,inner sep=1pt,shape=circle,color=gray] {};
\path (1.33333,2.) node(23_0) [fill,inner sep=1pt,shape=circle,color=gray] {};
\path (2.,2.) node(24_0) [fill,inner sep=1pt,shape=circle,color=gray] {};
\draw[color=gray] (9_0) -- (10_0);
\draw[color=gray] (10_0) -- (11_0);
\draw[color=gray] (11_0) -- (12_0);
\draw[color=gray] (13_0) -- (14_0);
\draw[color=gray] (14_0) -- (15_0);
\draw[color=gray] (15_0) -- (16_0);
\draw[color=gray] (17_0) -- (18_0);
\draw[color=gray] (18_0) -- (19_0);
\draw[color=gray] (19_0) -- (20_0);
\draw[color=gray] (21_0) -- (22_0);
\draw[color=gray] (22_0) -- (23_0);
\draw[color=gray] (23_0) -- (24_0);
\draw[color=gray] (9_0) -- (13_0);
\draw[color=gray] (13_0) -- (17_0);
\draw[color=gray] (17_0) -- (21_0);
\draw[color=gray] (10_0) -- (14_0);
\draw[color=gray] (14_0) -- (18_0);
\draw[color=gray] (18_0) -- (22_0);
\draw[color=gray] (11_0) -- (15_0);
\draw[color=gray] (15_0) -- (19_0);
\draw[color=gray] (19_0) -- (23_0);
\draw[color=gray] (12_0) -- (16_0);
\draw[color=gray] (16_0) -- (20_0);
\draw[color=gray] (20_0) -- (24_0);
\path (0.333333,0.333333) node(0_0) [] {};
\path (1.,0.333333) node(1_0) [] {};
\path (1.66667,0.333333) node(2_0) [] {};
\path (0.333333,1.) node(3_0) [] {};
\path (1.,1.) node(4_0) [] {};
\path (1.66667,1.) node(5_0) [] {};
\path (0.333333,1.66667) node(6_0) [] {};
\path (1.,1.66667) node(7_0) [] {};
\path (1.66667,1.66667) node(8_0) [] {};
\end{tikzpicture}
\end{center}
Now we can extrude this mesh 4 levels to produce a $3 \times 3 \times 4$ brick,
\begin{Verbatim}[fontsize=\small]
  ./ex1 -dm_plex_simplex 0 -dm_plex_box_faces 3,3
    -dm_extrude 4
    -dm_view :mesh.tex:ascii_latex
    -dm_plex_view_numbers 0
    -dm_plex_view_colors_depth 1,0,0,0
    -dm_plex_view_scale 4
\end{Verbatim}
shown below.
\begin{center}
\begin{tikzpicture}[scale = 1.,font=\fontsize{8}{8}\selectfont]
\path (0.,0.,-0.) node(36_0) [fill,inner sep=1pt,shape=circle,color=gray] {};
\path (0.,1.,-0.) node(37_0) [fill,inner sep=1pt,shape=circle,color=gray] {};
\path (0.,2.,-0.) node(38_0) [fill,inner sep=1pt,shape=circle,color=gray] {};
\path (0.,3.,-0.) node(39_0) [fill,inner sep=1pt,shape=circle,color=gray] {};
\path (0.,4.,-0.) node(40_0) [fill,inner sep=1pt,shape=circle,color=gray] {};
\path (1.33333,0.,-0.) node(41_0) [fill,inner sep=1pt,shape=circle,color=gray] {};
\path (1.33333,1.,-0.) node(42_0) [fill,inner sep=1pt,shape=circle,color=gray] {};
\path (1.33333,2.,-0.) node(43_0) [fill,inner sep=1pt,shape=circle,color=gray] {};
\path (1.33333,3.,-0.) node(44_0) [fill,inner sep=1pt,shape=circle,color=gray] {};
\path (1.33333,4.,-0.) node(45_0) [fill,inner sep=1pt,shape=circle,color=gray] {};
\path (2.66667,0.,-0.) node(46_0) [fill,inner sep=1pt,shape=circle,color=gray] {};
\path (2.66667,1.,-0.) node(47_0) [fill,inner sep=1pt,shape=circle,color=gray] {};
\path (2.66667,2.,-0.) node(48_0) [fill,inner sep=1pt,shape=circle,color=gray] {};
\path (2.66667,3.,-0.) node(49_0) [fill,inner sep=1pt,shape=circle,color=gray] {};
\path (2.66667,4.,-0.) node(50_0) [fill,inner sep=1pt,shape=circle,color=gray] {};
\path (4.,0.,-0.) node(51_0) [fill,inner sep=1pt,shape=circle,color=gray] {};
\path (4.,1.,-0.) node(52_0) [fill,inner sep=1pt,shape=circle,color=gray] {};
\path (4.,2.,-0.) node(53_0) [fill,inner sep=1pt,shape=circle,color=gray] {};
\path (4.,3.,-0.) node(54_0) [fill,inner sep=1pt,shape=circle,color=gray] {};
\path (4.,4.,-0.) node(55_0) [fill,inner sep=1pt,shape=circle,color=gray] {};
\path (0.,0.,-1.33333) node(56_0) [fill,inner sep=1pt,shape=circle,color=gray] {};
\path (0.,1.,-1.33333) node(57_0) [fill,inner sep=1pt,shape=circle,color=gray] {};
\path (0.,2.,-1.33333) node(58_0) [fill,inner sep=1pt,shape=circle,color=gray] {};
\path (0.,3.,-1.33333) node(59_0) [fill,inner sep=1pt,shape=circle,color=gray] {};
\path (0.,4.,-1.33333) node(60_0) [fill,inner sep=1pt,shape=circle,color=gray] {};
\path (1.33333,0.,-1.33333) node(61_0) [fill,inner sep=1pt,shape=circle,color=gray] {};
\path (1.33333,1.,-1.33333) node(62_0) [fill,inner sep=1pt,shape=circle,color=gray] {};
\path (1.33333,2.,-1.33333) node(63_0) [fill,inner sep=1pt,shape=circle,color=gray] {};
\path (1.33333,3.,-1.33333) node(64_0) [fill,inner sep=1pt,shape=circle,color=gray] {};
\path (1.33333,4.,-1.33333) node(65_0) [fill,inner sep=1pt,shape=circle,color=gray] {};
\path (2.66667,0.,-1.33333) node(66_0) [fill,inner sep=1pt,shape=circle,color=gray] {};
\path (2.66667,1.,-1.33333) node(67_0) [fill,inner sep=1pt,shape=circle,color=gray] {};
\path (2.66667,2.,-1.33333) node(68_0) [fill,inner sep=1pt,shape=circle,color=gray] {};
\path (2.66667,3.,-1.33333) node(69_0) [fill,inner sep=1pt,shape=circle,color=gray] {};
\path (2.66667,4.,-1.33333) node(70_0) [fill,inner sep=1pt,shape=circle,color=gray] {};
\path (4.,0.,-1.33333) node(71_0) [fill,inner sep=1pt,shape=circle,color=gray] {};
\path (4.,1.,-1.33333) node(72_0) [fill,inner sep=1pt,shape=circle,color=gray] {};
\path (4.,2.,-1.33333) node(73_0) [fill,inner sep=1pt,shape=circle,color=gray] {};
\path (4.,3.,-1.33333) node(74_0) [fill,inner sep=1pt,shape=circle,color=gray] {};
\path (4.,4.,-1.33333) node(75_0) [fill,inner sep=1pt,shape=circle,color=gray] {};
\path (0.,0.,-2.66667) node(76_0) [fill,inner sep=1pt,shape=circle,color=gray] {};
\path (0.,1.,-2.66667) node(77_0) [fill,inner sep=1pt,shape=circle,color=gray] {};
\path (0.,2.,-2.66667) node(78_0) [fill,inner sep=1pt,shape=circle,color=gray] {};
\path (0.,3.,-2.66667) node(79_0) [fill,inner sep=1pt,shape=circle,color=gray] {};
\path (0.,4.,-2.66667) node(80_0) [fill,inner sep=1pt,shape=circle,color=gray] {};
\path (1.33333,0.,-2.66667) node(81_0) [fill,inner sep=1pt,shape=circle,color=gray] {};
\path (1.33333,1.,-2.66667) node(82_0) [fill,inner sep=1pt,shape=circle,color=gray] {};
\path (1.33333,2.,-2.66667) node(83_0) [fill,inner sep=1pt,shape=circle,color=gray] {};
\path (1.33333,3.,-2.66667) node(84_0) [fill,inner sep=1pt,shape=circle,color=gray] {};
\path (1.33333,4.,-2.66667) node(85_0) [fill,inner sep=1pt,shape=circle,color=gray] {};
\path (2.66667,0.,-2.66667) node(86_0) [fill,inner sep=1pt,shape=circle,color=gray] {};
\path (2.66667,1.,-2.66667) node(87_0) [fill,inner sep=1pt,shape=circle,color=gray] {};
\path (2.66667,2.,-2.66667) node(88_0) [fill,inner sep=1pt,shape=circle,color=gray] {};
\path (2.66667,3.,-2.66667) node(89_0) [fill,inner sep=1pt,shape=circle,color=gray] {};
\path (2.66667,4.,-2.66667) node(90_0) [fill,inner sep=1pt,shape=circle,color=gray] {};
\path (4.,0.,-2.66667) node(91_0) [fill,inner sep=1pt,shape=circle,color=gray] {};
\path (4.,1.,-2.66667) node(92_0) [fill,inner sep=1pt,shape=circle,color=gray] {};
\path (4.,2.,-2.66667) node(93_0) [fill,inner sep=1pt,shape=circle,color=gray] {};
\path (4.,3.,-2.66667) node(94_0) [fill,inner sep=1pt,shape=circle,color=gray] {};
\path (4.,4.,-2.66667) node(95_0) [fill,inner sep=1pt,shape=circle,color=gray] {};
\path (0.,0.,-4.) node(96_0) [fill,inner sep=1pt,shape=circle,color=gray] {};
\path (0.,1.,-4.) node(97_0) [fill,inner sep=1pt,shape=circle,color=gray] {};
\path (0.,2.,-4.) node(98_0) [fill,inner sep=1pt,shape=circle,color=gray] {};
\path (0.,3.,-4.) node(99_0) [fill,inner sep=1pt,shape=circle,color=gray] {};
\path (0.,4.,-4.) node(100_0) [fill,inner sep=1pt,shape=circle,color=gray] {};
\path (1.33333,0.,-4.) node(101_0) [fill,inner sep=1pt,shape=circle,color=gray] {};
\path (1.33333,1.,-4.) node(102_0) [fill,inner sep=1pt,shape=circle,color=gray] {};
\path (1.33333,2.,-4.) node(103_0) [fill,inner sep=1pt,shape=circle,color=gray] {};
\path (1.33333,3.,-4.) node(104_0) [fill,inner sep=1pt,shape=circle,color=gray] {};
\path (1.33333,4.,-4.) node(105_0) [fill,inner sep=1pt,shape=circle,color=gray] {};
\path (2.66667,0.,-4.) node(106_0) [fill,inner sep=1pt,shape=circle,color=gray] {};
\path (2.66667,1.,-4.) node(107_0) [fill,inner sep=1pt,shape=circle,color=gray] {};
\path (2.66667,2.,-4.) node(108_0) [fill,inner sep=1pt,shape=circle,color=gray] {};
\path (2.66667,3.,-4.) node(109_0) [fill,inner sep=1pt,shape=circle,color=gray] {};
\path (2.66667,4.,-4.) node(110_0) [fill,inner sep=1pt,shape=circle,color=gray] {};
\path (4.,0.,-4.) node(111_0) [fill,inner sep=1pt,shape=circle,color=gray] {};
\path (4.,1.,-4.) node(112_0) [fill,inner sep=1pt,shape=circle,color=gray] {};
\path (4.,2.,-4.) node(113_0) [fill,inner sep=1pt,shape=circle,color=gray] {};
\path (4.,3.,-4.) node(114_0) [fill,inner sep=1pt,shape=circle,color=gray] {};
\path (4.,4.,-4.) node(115_0) [fill,inner sep=1pt,shape=circle,color=gray] {};
\draw[color=gray] (36_0) -- (41_0);
\draw[color=gray] (37_0) -- (42_0);
\draw[color=gray] (38_0) -- (43_0);
\draw[color=gray] (39_0) -- (44_0);
\draw[color=gray] (40_0) -- (45_0);
\draw[color=gray] (41_0) -- (46_0);
\draw[color=gray] (42_0) -- (47_0);
\draw[color=gray] (43_0) -- (48_0);
\draw[color=gray] (44_0) -- (49_0);
\draw[color=gray] (45_0) -- (50_0);
\draw[color=gray] (46_0) -- (51_0);
\draw[color=gray] (47_0) -- (52_0);
\draw[color=gray] (48_0) -- (53_0);
\draw[color=gray] (49_0) -- (54_0);
\draw[color=gray] (50_0) -- (55_0);
\draw[color=gray] (56_0) -- (61_0);
\draw[color=gray] (57_0) -- (62_0);
\draw[color=gray] (58_0) -- (63_0);
\draw[color=gray] (59_0) -- (64_0);
\draw[color=gray] (60_0) -- (65_0);
\draw[color=gray] (61_0) -- (66_0);
\draw[color=gray] (62_0) -- (67_0);
\draw[color=gray] (63_0) -- (68_0);
\draw[color=gray] (64_0) -- (69_0);
\draw[color=gray] (65_0) -- (70_0);
\draw[color=gray] (66_0) -- (71_0);
\draw[color=gray] (67_0) -- (72_0);
\draw[color=gray] (68_0) -- (73_0);
\draw[color=gray] (69_0) -- (74_0);
\draw[color=gray] (70_0) -- (75_0);
\draw[color=gray] (76_0) -- (81_0);
\draw[color=gray] (77_0) -- (82_0);
\draw[color=gray] (78_0) -- (83_0);
\draw[color=gray] (79_0) -- (84_0);
\draw[color=gray] (80_0) -- (85_0);
\draw[color=gray] (81_0) -- (86_0);
\draw[color=gray] (82_0) -- (87_0);
\draw[color=gray] (83_0) -- (88_0);
\draw[color=gray] (84_0) -- (89_0);
\draw[color=gray] (85_0) -- (90_0);
\draw[color=gray] (86_0) -- (91_0);
\draw[color=gray] (87_0) -- (92_0);
\draw[color=gray] (88_0) -- (93_0);
\draw[color=gray] (89_0) -- (94_0);
\draw[color=gray] (90_0) -- (95_0);
\draw[color=gray] (96_0) -- (101_0);
\draw[color=gray] (97_0) -- (102_0);
\draw[color=gray] (98_0) -- (103_0);
\draw[color=gray] (99_0) -- (104_0);
\draw[color=gray] (100_0) -- (105_0);
\draw[color=gray] (101_0) -- (106_0);
\draw[color=gray] (102_0) -- (107_0);
\draw[color=gray] (103_0) -- (108_0);
\draw[color=gray] (104_0) -- (109_0);
\draw[color=gray] (105_0) -- (110_0);
\draw[color=gray] (106_0) -- (111_0);
\draw[color=gray] (107_0) -- (112_0);
\draw[color=gray] (108_0) -- (113_0);
\draw[color=gray] (109_0) -- (114_0);
\draw[color=gray] (110_0) -- (115_0);
\draw[color=gray] (36_0) -- (56_0);
\draw[color=gray] (37_0) -- (57_0);
\draw[color=gray] (38_0) -- (58_0);
\draw[color=gray] (39_0) -- (59_0);
\draw[color=gray] (40_0) -- (60_0);
\draw[color=gray] (56_0) -- (76_0);
\draw[color=gray] (57_0) -- (77_0);
\draw[color=gray] (58_0) -- (78_0);
\draw[color=gray] (59_0) -- (79_0);
\draw[color=gray] (60_0) -- (80_0);
\draw[color=gray] (76_0) -- (96_0);
\draw[color=gray] (77_0) -- (97_0);
\draw[color=gray] (78_0) -- (98_0);
\draw[color=gray] (79_0) -- (99_0);
\draw[color=gray] (80_0) -- (100_0);
\draw[color=gray] (41_0) -- (61_0);
\draw[color=gray] (42_0) -- (62_0);
\draw[color=gray] (43_0) -- (63_0);
\draw[color=gray] (44_0) -- (64_0);
\draw[color=gray] (45_0) -- (65_0);
\draw[color=gray] (61_0) -- (81_0);
\draw[color=gray] (62_0) -- (82_0);
\draw[color=gray] (63_0) -- (83_0);
\draw[color=gray] (64_0) -- (84_0);
\draw[color=gray] (65_0) -- (85_0);
\draw[color=gray] (81_0) -- (101_0);
\draw[color=gray] (82_0) -- (102_0);
\draw[color=gray] (83_0) -- (103_0);
\draw[color=gray] (84_0) -- (104_0);
\draw[color=gray] (85_0) -- (105_0);
\draw[color=gray] (46_0) -- (66_0);
\draw[color=gray] (47_0) -- (67_0);
\draw[color=gray] (48_0) -- (68_0);
\draw[color=gray] (49_0) -- (69_0);
\draw[color=gray] (50_0) -- (70_0);
\draw[color=gray] (66_0) -- (86_0);
\draw[color=gray] (67_0) -- (87_0);
\draw[color=gray] (68_0) -- (88_0);
\draw[color=gray] (69_0) -- (89_0);
\draw[color=gray] (70_0) -- (90_0);
\draw[color=gray] (86_0) -- (106_0);
\draw[color=gray] (87_0) -- (107_0);
\draw[color=gray] (88_0) -- (108_0);
\draw[color=gray] (89_0) -- (109_0);
\draw[color=gray] (90_0) -- (110_0);
\draw[color=gray] (51_0) -- (71_0);
\draw[color=gray] (52_0) -- (72_0);
\draw[color=gray] (53_0) -- (73_0);
\draw[color=gray] (54_0) -- (74_0);
\draw[color=gray] (55_0) -- (75_0);
\draw[color=gray] (71_0) -- (91_0);
\draw[color=gray] (72_0) -- (92_0);
\draw[color=gray] (73_0) -- (93_0);
\draw[color=gray] (74_0) -- (94_0);
\draw[color=gray] (75_0) -- (95_0);
\draw[color=gray] (91_0) -- (111_0);
\draw[color=gray] (92_0) -- (112_0);
\draw[color=gray] (93_0) -- (113_0);
\draw[color=gray] (94_0) -- (114_0);
\draw[color=gray] (95_0) -- (115_0);
\draw[color=gray] (36_0) -- (37_0);
\draw[color=gray] (37_0) -- (38_0);
\draw[color=gray] (38_0) -- (39_0);
\draw[color=gray] (39_0) -- (40_0);
\draw[color=gray] (41_0) -- (42_0);
\draw[color=gray] (42_0) -- (43_0);
\draw[color=gray] (43_0) -- (44_0);
\draw[color=gray] (44_0) -- (45_0);
\draw[color=gray] (46_0) -- (47_0);
\draw[color=gray] (47_0) -- (48_0);
\draw[color=gray] (48_0) -- (49_0);
\draw[color=gray] (49_0) -- (50_0);
\draw[color=gray] (51_0) -- (52_0);
\draw[color=gray] (52_0) -- (53_0);
\draw[color=gray] (53_0) -- (54_0);
\draw[color=gray] (54_0) -- (55_0);
\draw[color=gray] (56_0) -- (57_0);
\draw[color=gray] (57_0) -- (58_0);
\draw[color=gray] (58_0) -- (59_0);
\draw[color=gray] (59_0) -- (60_0);
\draw[color=gray] (61_0) -- (62_0);
\draw[color=gray] (62_0) -- (63_0);
\draw[color=gray] (63_0) -- (64_0);
\draw[color=gray] (64_0) -- (65_0);
\draw[color=gray] (66_0) -- (67_0);
\draw[color=gray] (67_0) -- (68_0);
\draw[color=gray] (68_0) -- (69_0);
\draw[color=gray] (69_0) -- (70_0);
\draw[color=gray] (71_0) -- (72_0);
\draw[color=gray] (72_0) -- (73_0);
\draw[color=gray] (73_0) -- (74_0);
\draw[color=gray] (74_0) -- (75_0);
\draw[color=gray] (76_0) -- (77_0);
\draw[color=gray] (77_0) -- (78_0);
\draw[color=gray] (78_0) -- (79_0);
\draw[color=gray] (79_0) -- (80_0);
\draw[color=gray] (81_0) -- (82_0);
\draw[color=gray] (82_0) -- (83_0);
\draw[color=gray] (83_0) -- (84_0);
\draw[color=gray] (84_0) -- (85_0);
\draw[color=gray] (86_0) -- (87_0);
\draw[color=gray] (87_0) -- (88_0);
\draw[color=gray] (88_0) -- (89_0);
\draw[color=gray] (89_0) -- (90_0);
\draw[color=gray] (91_0) -- (92_0);
\draw[color=gray] (92_0) -- (93_0);
\draw[color=gray] (93_0) -- (94_0);
\draw[color=gray] (94_0) -- (95_0);
\draw[color=gray] (96_0) -- (97_0);
\draw[color=gray] (97_0) -- (98_0);
\draw[color=gray] (98_0) -- (99_0);
\draw[color=gray] (99_0) -- (100_0);
\draw[color=gray] (101_0) -- (102_0);
\draw[color=gray] (102_0) -- (103_0);
\draw[color=gray] (103_0) -- (104_0);
\draw[color=gray] (104_0) -- (105_0);
\draw[color=gray] (106_0) -- (107_0);
\draw[color=gray] (107_0) -- (108_0);
\draw[color=gray] (108_0) -- (109_0);
\draw[color=gray] (109_0) -- (110_0);
\draw[color=gray] (111_0) -- (112_0);
\draw[color=gray] (112_0) -- (113_0);
\draw[color=gray] (113_0) -- (114_0);
\draw[color=gray] (114_0) -- (115_0);
\path (0.666667,0.5,-0.666667) node(0_0) [] {};
\path (0.666667,1.5,-0.666667) node(1_0) [] {};
\path (0.666667,2.5,-0.666667) node(2_0) [] {};
\path (0.666667,3.5,-0.666667) node(3_0) [] {};
\path (2.,0.5,-0.666667) node(4_0) [] {};
\path (2.,1.5,-0.666667) node(5_0) [] {};
\path (2.,2.5,-0.666667) node(6_0) [] {};
\path (2.,3.5,-0.666667) node(7_0) [] {};
\path (3.33333,0.5,-0.666667) node(8_0) [] {};
\path (3.33333,1.5,-0.666667) node(9_0) [] {};
\path (3.33333,2.5,-0.666667) node(10_0) [] {};
\path (3.33333,3.5,-0.666667) node(11_0) [] {};
\path (0.666667,0.5,-2.) node(12_0) [] {};
\path (0.666667,1.5,-2.) node(13_0) [] {};
\path (0.666667,2.5,-2.) node(14_0) [] {};
\path (0.666667,3.5,-2.) node(15_0) [] {};
\path (2.,0.5,-2.) node(16_0) [] {};
\path (2.,1.5,-2.) node(17_0) [] {};
\path (2.,2.5,-2.) node(18_0) [] {};
\path (2.,3.5,-2.) node(19_0) [] {};
\path (3.33333,0.5,-2.) node(20_0) [] {};
\path (3.33333,1.5,-2.) node(21_0) [] {};
\path (3.33333,2.5,-2.) node(22_0) [] {};
\path (3.33333,3.5,-2.) node(23_0) [] {};
\path (0.666667,0.5,-3.33333) node(24_0) [] {};
\path (0.666667,1.5,-3.33333) node(25_0) [] {};
\path (0.666667,2.5,-3.33333) node(26_0) [] {};
\path (0.666667,3.5,-3.33333) node(27_0) [] {};
\path (2.,0.5,-3.33333) node(28_0) [] {};
\path (2.,1.5,-3.33333) node(29_0) [] {};
\path (2.,2.5,-3.33333) node(30_0) [] {};
\path (2.,3.5,-3.33333) node(31_0) [] {};
\path (3.33333,0.5,-3.33333) node(32_0) [] {};
\path (3.33333,1.5,-3.33333) node(33_0) [] {};
\path (3.33333,2.5,-3.33333) node(34_0) [] {};
\path (3.33333,3.5,-3.33333) node(35_0) [] {};
\end{tikzpicture}
\end{center}

\subsection{Simple Extrusion}

When automating extrusion, we would like to allow for an arbitrary number of layers. Thus we need dynamic data structures to indicate how cells transform, rather than a static definition. We use a structure with five parts to hold our information,
\begin{enumerate}
  \item {\bf Nt}: the number of celltypes produced
  \item {\bf target}: the celltypes produced
  \item {\bf size}: the number of each type of cells
  \item {\bf cone}: the cone of each produced cells
  \item {\bf ornt}: the orientation of each cone point
\end{enumerate}
The \cinline{cone} array is encoded using the scheme we laid out in Section~\ref{sec:transformDef}, namely that each entry has a celltype, the number of cones to be taken, the cone index for each, and finally the replica number. In order to describe extrusion for all cells, we index this structure by celltype. As the simplest example, below we explain how a vertex is extruded into a series of line segments.

We first indicate that this description is for celltype {\sc point}, and then we set the number of celltypes produced to two. We make more vertices, and then either {\sc segment}s or {\sc point\_prism\_tensor}s, depending on whether we decide to use tensor cells in the extrusion. We use the number of layers, \cinline{Nl}, to allocate the rest of our data structure.
\begin{cprog}
ct = DM_POLYTOPE_POINT;
Nt[ct] = 2;
Nc = 6*Nl;
No = 2*Nl;
PetscMalloc4(Nt[ct], &target[ct], Nt[ct], &size[ct],
             Nc, &cone[ct], No, &ornt[ct]);
target[ct][0] = DM_POLYTOPE_POINT;
target[ct][1] = useTensor ? DM_POLYTOPE_POINT_PRISM_TENSOR
                : DM_POLYTOPE_SEGMENT;
size[ct][0]   = Nl+1;
size[ct][1]   = Nl;
for (i = 0; i < Nl; ++i) {
  cone[ct][6*i+0] = DM_POLYTOPE_POINT;
  cone[ct][6*i+1] = 0;
  cone[ct][6*i+2] = i;
  cone[ct][6*i+3] = DM_POLYTOPE_POINT;
  cone[ct][6*i+4] = 0;
  cone[ct][6*i+5] = i+1;
}
for (i = 0; i < No; ++i) ornt[ct][i] = 0;
\end{cprog}
For \cinline{Nl} layers, we have \cinline{Nl} segments and \cinline{Nl+1} vertices. The vertices have no cones, but for each segment, we have two vertices which are described solely by replica number since they are all produced by the original vertex. Finally, vertices always have orientation 0.

A less trivial example is provided by the extrusion of triangles. We again produce two celltypes, {\sc triangle}s and either {\sc tri\_prism}s or {\sc tri\_prism\_tensor}s, and we must specify cones and orientations for both. As before, we make \cinline{Nl} prisms and \cinline{Nl+1} triangles.
\begin{cprog}
ct = DM_POLYTOPE_TRIANGLE;
Nt[ct] = 2;
Nc = 12*(Nl+1) + 18*Nl;
No =  3*(Nl+1) +  5*Nl;
PetscMalloc4(Nt[ct], &target[ct], Nt[ct], &size[ct],
             Nc, &cone[ct], No, &ornt[ct]);
target[ct][0] = DM_POLYTOPE_TRIANGLE;
target[ct][1] = useTensor ? DM_POLYTOPE_TRI_PRISM_TENSOR
                : DM_POLYTOPE_TRI_PRISM;
size[ct][0]   = Nl+1;
size[ct][1]   = Nl;
\end{cprog}
The cones for the triangles are straightforward. They are each formed from the edges produced by the edges of the original triangle, and in the same order. Thus, the replica numbers are exactly the layer number, and the orientations are zero. If the orientation of the original triangle is nonzero, this is propagated by the group action mentioned above.
\begin{cprog}
for (i = 0; i < Nl+1; ++i) {
  cone[ct][12*i+0]  = DM_POLYTOPE_SEGMENT;
  cone[ct][12*i+1]  = 1;
  cone[ct][12*i+2]  = 0;
  cone[ct][12*i+3]  = i;
  cone[ct][12*i+4]  = DM_POLYTOPE_SEGMENT;
  cone[ct][12*i+5]  = 1;
  cone[ct][12*i+6]  = 1;
  cone[ct][12*i+7]  = i;
  cone[ct][12*i+8]  = DM_POLYTOPE_SEGMENT;
  cone[ct][12*i+9]  = 1;
  cone[ct][12*i+10] = 2;
  cone[ct][12*i+11] = i;
}
for (i = 0; i < 3*(Nl+1); ++i) ornt[ct][i] = 0;
\end{cprog}
Finally, we construct the cones of the triangular prisms. Each consists of two triangle endcaps, and three side faces which can be either {\sc seg\_prism\_tensor}s or {\sc quadrilateral}s. The side faces are those extruded by the edges of the original triangle, and are all properly oriented. However, if we make {\sc tri\_prism}s then the bottom endcap must reverse orientation so that it has an outward normal.
\begin{cprog}
coff = 12*(Nl+1);
ooff = 3*(Nl+1);
for (i = 0; i < Nl; ++i) {
  if (useTensor) {
    cone[ct][coff+18*i+0]  = DM_POLYTOPE_TRIANGLE;
    cone[ct][coff+18*i+1]  = 0;
    cone[ct][coff+18*i+2]  = i;
    cone[ct][coff+18*i+3]  = DM_POLYTOPE_TRIANGLE;
    cone[ct][coff+18*i+4]  = 0;
    cone[ct][coff+18*i+5]  = i+1;
    cone[ct][coff+18*i+6]  = DM_POLYTOPE_SEG_PRISM_TENSOR;
    cone[ct][coff+18*i+7]  = 1;
    cone[ct][coff+18*i+8]  = 0;
    cone[ct][coff+18*i+9]  = i;
    cone[ct][coff+18*i+10] = DM_POLYTOPE_SEG_PRISM_TENSOR;
    cone[ct][coff+18*i+11] = 1;
    cone[ct][coff+18*i+12] = 1;
    cone[ct][coff+18*i+13] = i;
    cone[ct][coff+18*i+14] = DM_POLYTOPE_SEG_PRISM_TENSOR;
    cone[ct][coff+18*i+15] = 1;
    cone[ct][coff+18*i+16] = 2;
    cone[ct][coff+18*i+17] = i;
    ornt[ct][ooff+5*i+0] = 0;
    ornt[ct][ooff+5*i+1] = 0;
    ornt[ct][ooff+5*i+2] = 0;
    ornt[ct][ooff+5*i+3] = 0;
    ornt[ct][ooff+5*i+4] = 0;
  } else {
    cone[ct][coff+18*i+0]  = DM_POLYTOPE_TRIANGLE;
    cone[ct][coff+18*i+1]  = 0;
    cone[ct][coff+18*i+2]  = i;
    cone[ct][coff+18*i+3]  = DM_POLYTOPE_TRIANGLE;
    cone[ct][coff+18*i+4]  = 0;
    cone[ct][coff+18*i+5]  = i+1;
    cone[ct][coff+18*i+6]  = DM_POLYTOPE_QUADRILATERAL;
    cone[ct][coff+18*i+7]  = 1;
    cone[ct][coff+18*i+8]  = 0;
    cone[ct][coff+18*i+9]  = i;
    cone[ct][coff+18*i+10] = DM_POLYTOPE_QUADRILATERAL;
    cone[ct][coff+18*i+11] = 1;
    cone[ct][coff+18*i+12] = 1;
    cone[ct][coff+18*i+13] = i;
    cone[ct][coff+18*i+14] = DM_POLYTOPE_QUADRILATERAL;
    cone[ct][coff+18*i+15] = 1;
    cone[ct][coff+18*i+16] = 2;
    cone[ct][coff+18*i+17] = i;
    ornt[ct][ooff+5*i+0] = -2;
    ornt[ct][ooff+5*i+1] =  0;
    ornt[ct][ooff+5*i+2] =  0;
    ornt[ct][ooff+5*i+3] =  0;
    ornt[ct][ooff+5*i+4] =  0;
  }
}
\end{cprog}

Points produced in layers are numbered consecutively, and thus it is possible to segregate the points in each layer. For example, we can construct a label over all points whose value is the layer index. This would allow us to assemble equations which depended on layer index rather than normal coordinate, or to impose conditions related to the layer index.

\begin{figure}
\centering
\includegraphics[width=0.45\textwidth]{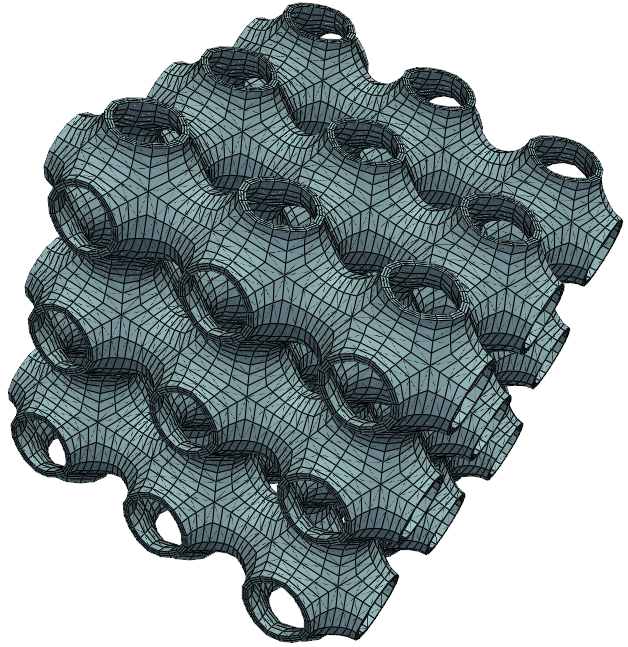}
\caption{The simplest Schwarz-P surface, extruded symmetrically about the function.\label{fig:tps}}
\end{figure}

\subsection{Coordinates}

New coordinates for the extruded mesh are determined by following the local normal out some distance from the original surface, and thus can be broken into two parts: computation of the local normal and computation of the layer start and thickness. We shall first look at determining the local normal direction. In the simplest case, we can prescribe a global normal direction, either using the \Verb+DMPlexTransformExtrudeSetNormal()+ method, or using a command line option, \Verb+-dm_plex_transform_extrude_normal+. If the normal is not specified and we are extruding an embedded surface, meaning the coordinates come from a higher dimensional space, then we can compute a local normal. The local normal at a mesh point $p$ is defined to be the average of the cell normals for all cells contained in its star. It would be possible to weight this average, for example by the area of each cell, but this is not currently done. Finally, if the normal is not computed or specified, the default normal for a two-dimensional extruded mesh is $\vu{y}$, and for a three-dimensional mesh is $\vu{z}$. In addition, we can specify a function which transforms the initial normal, given the coordinate. For example, we use this to construct accurate normals for triply periodic surfaces for which we have analytic expressions, as show in \cref{fig:tps}.
\begin{Verbatim}[fontsize=\small]
  ./ex1 -dm_plex_shape schwarz_p
    -dm_plex_tps_extent 3,3,3
    -dm_plex_tps_refine 3 -dm_plex_tps_layers 3
    -dm_plex_tps_thickness 0.1
\end{Verbatim}
In parallel, we want the average normal to replicate the serial case. Therefore, we construct a vector field for the normal over the mesh, compute the local part, sum in the normals from the shared region using a local-to-global field communication, \Verb|DMLocalToGlobal| in PETSc, and then normalize the vector.

\subsection{Surfaces}

\begin{figure}
\centering
\includegraphics[width=0.4\textwidth]{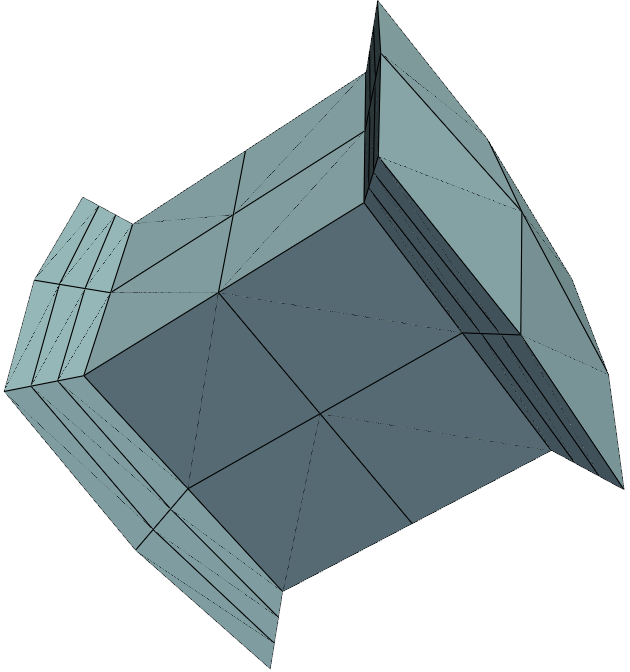}
\caption{Extrusion from the boundary of a cube.\label{fig:hexBdExt}}
\end{figure}

We can extrude only from a surface, for example to generate boundary layers. This surface can also be embedded, for example to generate cohesive cells for fracture mechanics and damage simulations. The extrusion surfaces are marked using the \textit{active} label for the transformation. In the example below, we use the boundary label for a cube, selecting two opposite faces, and extrude three layers
\begin{Verbatim}[fontsize=\small]
  ./ex1 -dm_plex_dim 3 -dm_plex_simplex 0 -dm_refine 1
    -dm_plex_box_faces 2,2,2 -dm_plex_separate_marker
    -dm_plex_transform_active marker
    -dm_plex_transform_active_values 1,2
    -dm_plex_transform_type extrude
    -dm_plex_transform_extrude_layers 3
    -dm_plex_transform_extrude_use_tensor 0
    -dm_plex_transform_extrude_thickness 0.25
\end{Verbatim}
giving us~\cref{fig:hexBdExt}. We can see that the normals are formed from the average of the normals in the star of each vertex.

\subsection{Refinement}

\begin{figure}
\centering
\includegraphics[width=0.35\textwidth]{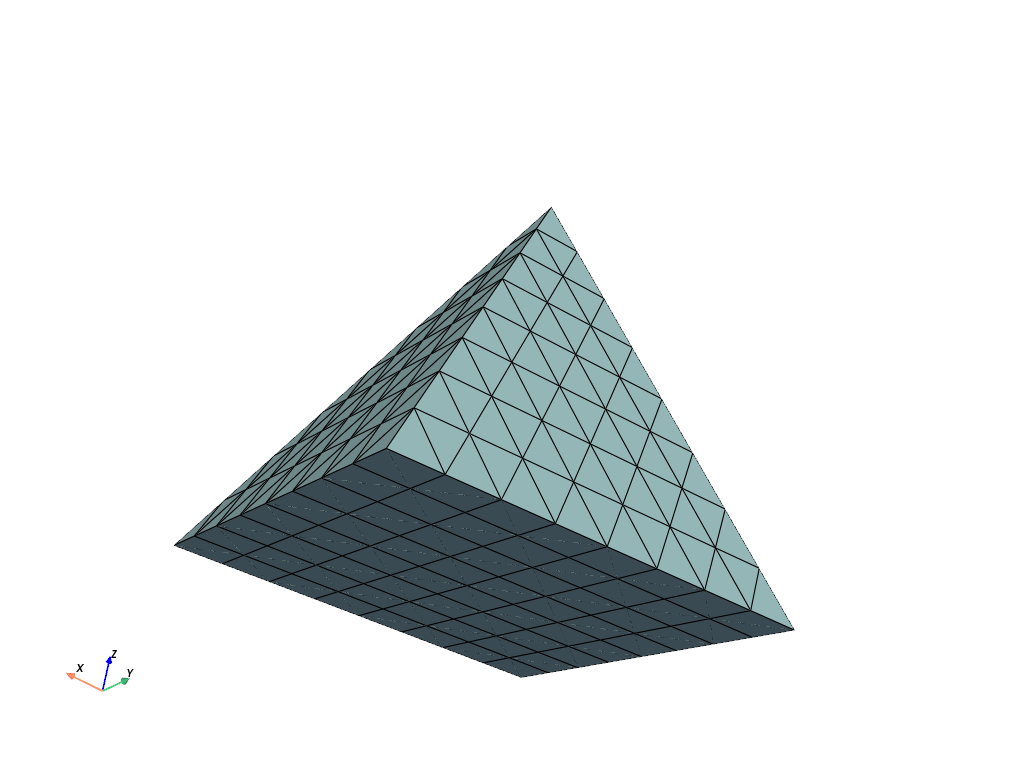}
\caption{Regular refinement of a pyramid.\label{fig:pyramidRef}}
\end{figure}

We showed above how rules can be written to regularly refine any cell type, for example a pyramid as shown in~\cref{fig:pyramidRef}.
\begin{Verbatim}[fontsize=\small]
  ./ex1 -dm_plex_dim 3 -dm_plex_reference_cell_domain
    -dm_plex_cell pyramid -dm_refine 3
\end{Verbatim}

\begin{figure}
\centering
\includegraphics[width=0.48\textwidth]{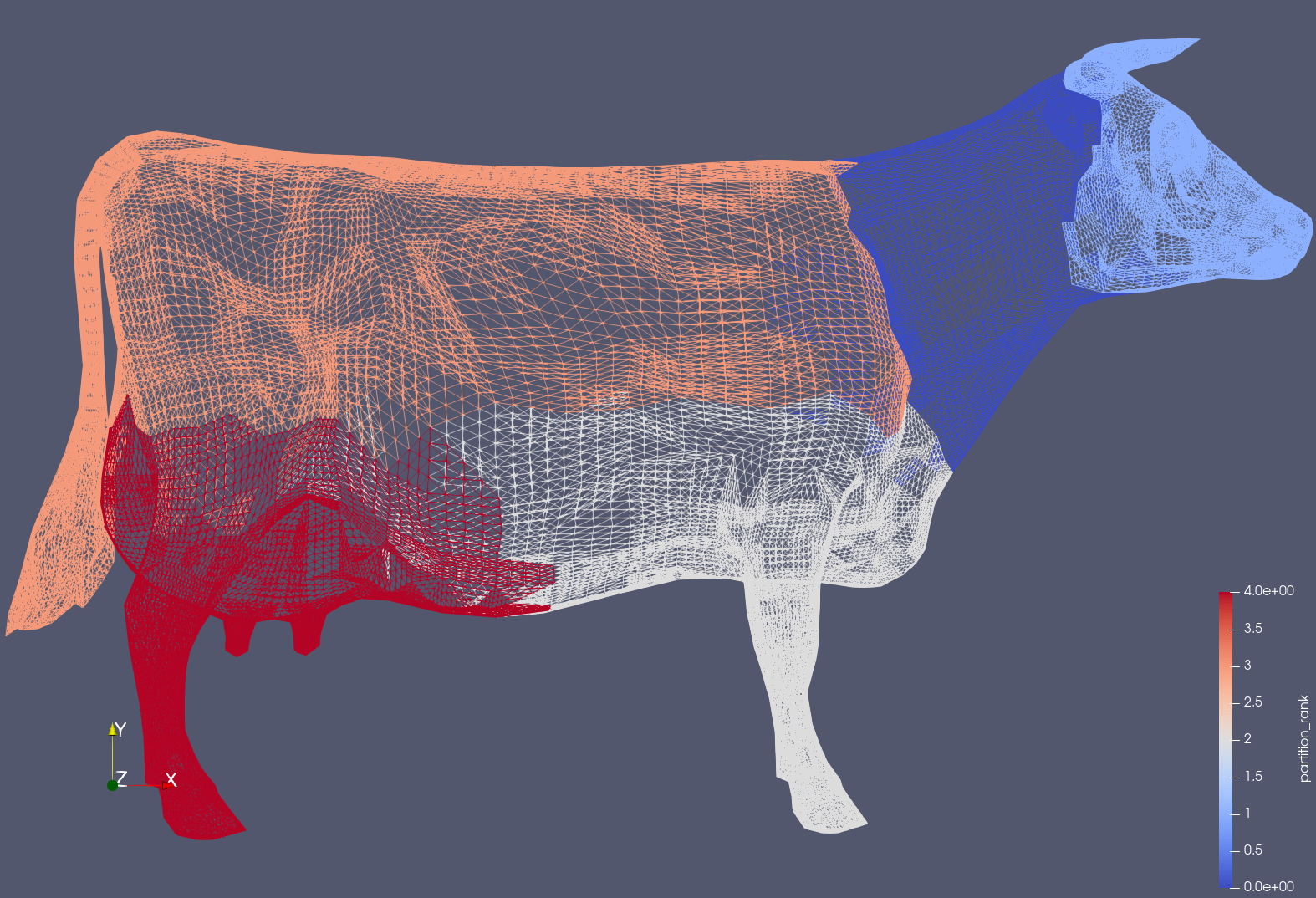}
\caption{Parallel refinement of an input mesh.\label{fig:cowRef}}
\end{figure}

This strategy can apply to mesh obtained from other sources, and work seamlessly in parallel, as in~\cref{fig:cowRef}.
\begin{Verbatim}[fontsize=\small]
  mpiexec -n 5 ./ex1 -dm_refine 2 -dm_plex_filename
    ${PETSC_DIR}/share/petsc/datafiles/meshes/cow.msh
    -dm_partition_view -dm_view hdf5:mesh.h5
    -dm_plex_view_hdf5_storage_version 1.1.0
\end{Verbatim}

\begin{figure}
\centering
\includegraphics[width=0.48\textwidth]{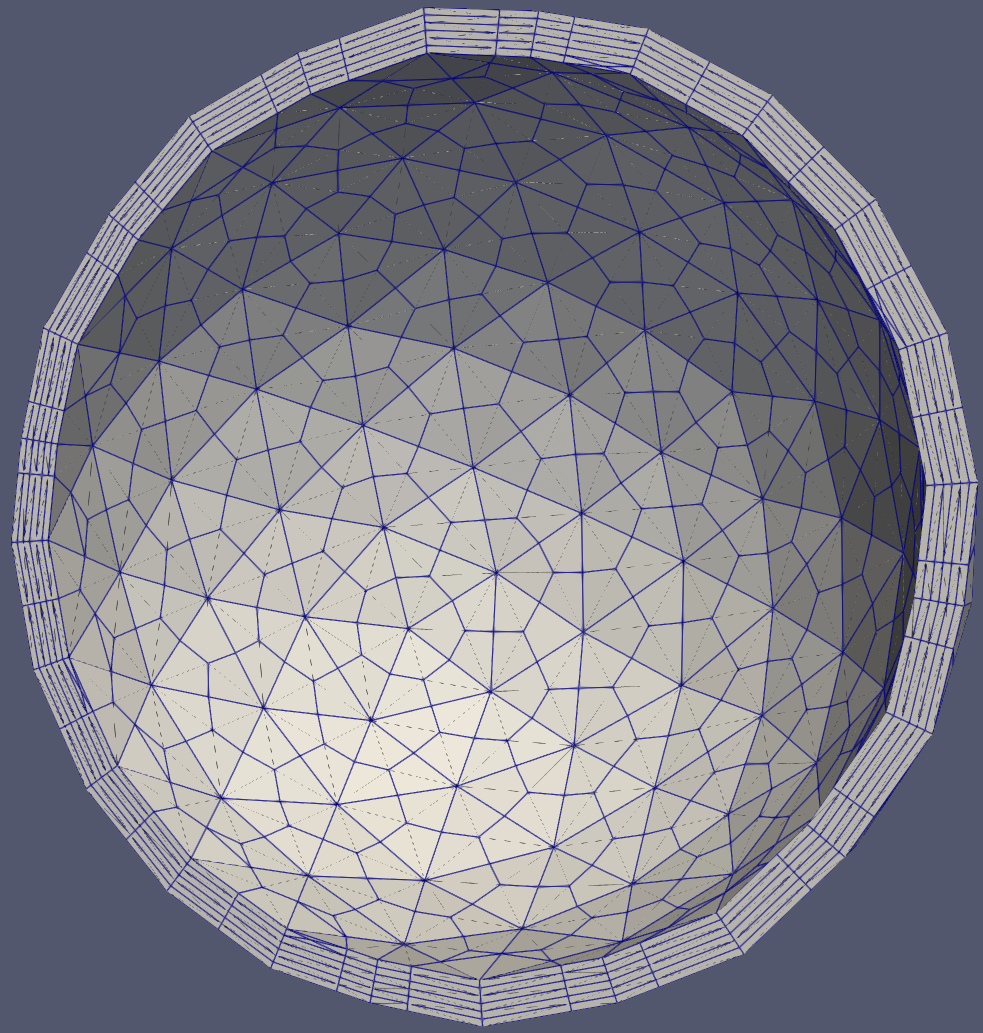}
\caption{Cut view of an extruded spherical shell converted to hexahedral cells.\label{fig:shellQuad}}
\end{figure}

We can also combine these transformations together. For example, we can generate a coarse representation of the 2-sphere, refine it, extrude the refined sphere, and then change the triangles to quadrilaterals, as shown in~\cref{fig:shellQuad}. Here we have used options prefixes in order to segregate processing into phases. This allows fine-grained control of the transformations from the command line, so that the code itself does not change.
\begin{Verbatim}[fontsize=\small]
  ./ex1 -dm_plex_option_phases ext_,box_
    -dm_plex_shape sphere -dm_refine 2
    -ext_dm_extrude 3
      -ext_dm_plex_transform_extrude_use_tensor 0
      -ext_dm_plex_transform_extrude_thickness 0.1
    -box_dm_refine 1
      -box_dm_plex_transform_type refine_tobox
    -dm_view hdf5:mesh.h5
\end{Verbatim}

\begin{figure}
\centering
\includegraphics[width=0.48\textwidth]{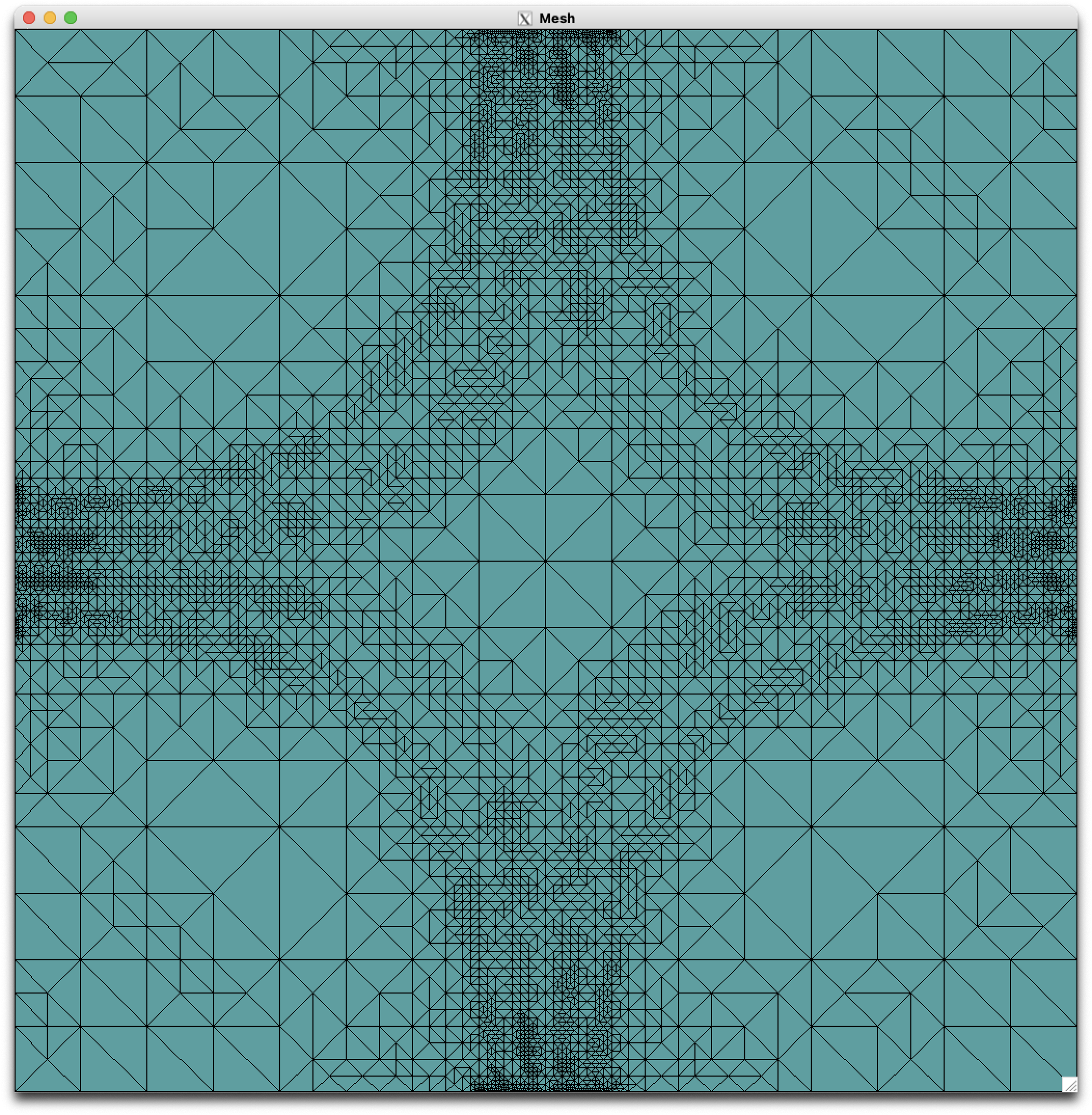}
\caption{Adaptively refined mesh for the Poisson equation based on flux-norm error estimates.\label{fig:SNESex27ensorAMR}}
\end{figure}

Using the ability to restrict or guide a transformation by labeling the mesh, we can adaptively refine with the Skeleton-Based Refinement algorithm of Plaza and Carey~\cite{plazacarey2000}. In PETSc SNES ex27, we solve the Poisson equation with a forcing function common in the AMR community
\begin{align}
  f(x, y) = \begin{cases} \sin(50 x y)/100 & \text{if}\ |xy| > 2\pi/50\\ \sin(50xy) & \text{otherwise} \end{cases}
\end{align}
We then estimate the error in the flux norm by solving the mixed form of Poisson, following the methodology outlined in~\cite{chamoinlegoll2023}. We use the PETSc VecTagger object to execute D\"orfler marking in order to refine the r\% of cells with the highest error estimator (here we use ten percent). The run can be executed using the PETSc test system to simplify the process, and the final mesh is shown in~\cref{fig:SNESex27ensorAMR}.
\begin{Verbatim}[fontsize=\small]
  make -f ./gmakefile test
    search="snes_tutorials-ex27_2d_p2_rt0p0_sensor"
    EXTRA_OPTIONS="-dm_adapt_pre_view draw
     -dm_adapt_iter_view draw -dm_adapt_view draw
     -snes_adapt_sequence 10 -dm_view_draw_cell_color 26"
\end{Verbatim}

\subsection{Ephemeral Meshes}\label{sec:ephemeral}

Notice that once we have a base mesh, as a concrete DMPlex object, and the table for a given transformation, we have enough information to answer any query about the transformed mesh. In fact, we need never concretely form the transformed mesh. Instead, we can keep the pair of base mesh and transform, and then run the transform machinery for each mesh query, such as a cone or closure. We call this pair an \textit{ephemeral mesh}. These can be created in PETSc using \Verb|DMPlexCreateEphemeral()| which takes as input a transform and the attached base mesh, and returns a valid DM object just as all other mesh creation routines.

As we have shown in~\cref{sec:grammar}, the complexity for computing any mesh query is \textit{output sensitive}, meaning the cost is proportional to the size of the output, rather than to the size of the entire original or transformed mesh. Thus, we may use ephemeral mesh anywhere we use a concrete mesh. In fact, we can compose ephemeral meshes using multiple transformations.

Consider the example of homogenization methods, such as the Localized Orthogonal Decomposition~\cite{maalqvistpeterseim2020}. A common feature is that a series of local problems must be solved on refined meshes overlaying the halo of a given coarse cell. To form these subproblems, we can filter the mesh to include a coarse cell and $n$ layers of overlap, and regularly refine this patch $k$ times. To do this concretely would likely involve many memory allocations and considerable churn on the heap. Instead we can create an emphemeral mesh for each patch, which merely answers queries based upon the specification. Of course, this logic could be written by hand, querying the original mesh. However, in the present form, it is composable, and it reuses all the code based upon the original PETSc DM interface for meshes.

\section{Conclusion}

We have given a formulation of mesh transformations as graph transformations defined by a restricted grammar which enables us to give both a compact representation and complexity guarantees for computations. These guarantees then enable us to construct ephemeral meshes, namely meshes which are not concretely instantiated, but rather mimiced by a base mesh/transformation pair. Our construction applies to a wide range of transformations, which we illustrate with examples, all of which come from the implementation in the PETSc library.

It seems clear that the tables assembled for each transformation should instead arise from some more compact mathematical description. Some initial work in this direction has already taken place. In~\cite{isaacknepley2017}, Toby Isaac proposes using a CW-complex to encode the mesh point transformations, and Dan Shapero in~\cite{shapero2025} encodes this information using a chain complex. However, in order to derive the parent-child orientation map, we need something like a chain complex with coefficients in the dihedral group of the faces, rather than $\mathbbm{Z}$. Perhaps the Isaac complex could have edges decorated with group members. This will be explored in future work.

\section*{Acknowledgements}
\anon
{Redacted for anonymity.} 
{ 
I thank all my gracious colleagues that contributed time, ideas, and hard work to make DMPlex a better package, including Brad Aagaard, Mark Adams, Nicolas Barral, Blaise Bourdin, Jed Brown, Carsten Burstedde, Lisandro Dalcin, Patrick Farrel, Gerard Gorman, David Ham, Toby Isaac, Michael Lange, Dmitry Karpeev, Rob Kirby, Dave May, Ridgway Scott, Barry Smith, Andy Terrel, Charles Williams, and Stefano Zampini.

This work was partially supported by DOE Applied Math Research, U.S. DOE Contract DE-AC02-06CH11357, and also Capturing the Dynamics of Compound Flooding in E3SM, U.S. DOE Contract DE-AC05-76RL01830.
}

\bibliographystyle{siam}
\bibliography{references}

\end{document}